\documentclass[12pt]{article}
\begin{document}

\begin{flushright}
RU--97-05\\
hep-th/yymmxxx\\
\today \\
\end{flushright}

\begin{center}

\vspace{0.5cm}
{\Large \bf Some Classical Solutions of Membrane \\ Matrix Model
Equations}
\vspace{0.5in}

{\bf Jens Hoppe}

\vspace{0.2in}

{\baselineskip=14pt
Institute for Theoretical Physics,\\ ETH H{\"o}nggerberg
CH8093, Z{\"u}rich, Switzerland}
\vspace{0.5cm}

{\baselineskip=14pt
Department of Physics and Astronomy, Rutgers University,\\
Piscataway, New Jersey 08855-0849, USA}
\vspace{1.5cm}

{\bf Abstract}
\end{center}

Some exact solutions to the classical matrix model equations that
arise in the context of M(embrane) theory are given, and their
topological nature is identified.

\newpage

Let $X_i(t)$, $i=1,...,d$, be time-dependent elements of some
(Lie-) algebra ${\cal A}$, satisfying (cp. \cite {1,2,3,4})

\begin{equation}
{\ddot X_i}=-\sum_{j=1}^{d}\bigl [[X_i,X_j],X_j\bigr ]
\end{equation}

\begin{equation}
\sum_{i=1}^{d}\bigl [X_i,{\dot X_i} \bigr ]=0.
\end{equation}

The Ansatz 
\begin{equation}
X_i(t)=x(t) r_{ij}(t)M_j
\end{equation}
(with $(r_{ij})_{ij=1...d}=e^{\bf A(t)}~\in ~SO(d)$ a time dependent
rotation, $x(t)$ an overall pulsation, ${\bf A}(t)=\varphi(t){\bf A}$,
${\bf A}^2 {\buildrel \rightarrow \over  M}=-\mu 
{\buildrel \rightarrow \over  M}$, $x^2(t){\dot \varphi (t)}=L=const.$)
reduces (1) to the equation(s)

\begin{equation}
\sum_{j=1}^d\bigl [[M_i,M_j],M_j\bigr ]=\lambda M_i,
\end{equation}
while $x(t)$ is then given as a solution of ${\ddot x}+\lambda x^3-
{\mu L^2\over x^2}=0$, i.e.

\begin{equation}
{1\over 2}{\dot x}^2+{\lambda \over 4}x^4+{\mu L^2\over 2}{1\over x^2}=E~~~;
\end{equation}

for $L\not= 0$ (when $L=0$, (2) automatically holds),
(2) may e.g. be satisfied by choosing half (or more) of the components
of ${ \buildrel \rightarrow \over  M}=( N_1,N_2,...
N_{\buildrel d\over \sim}, 0,...,0)$ to
vanish,
and ${\bf A}_{ij}$ to vanish whenever $M_i$ and $M_j$ don't, such as in

\begin{equation}
X_i=x(t)(cos\varphi(t){\buildrel \rightarrow \over   N},
sin\varphi(t){\buildrel \rightarrow \over  N}
, 0...0).
\end{equation}
Equation (4), which (for ${\buildrel \rightarrow \over  M}^2={\bf 1})$ may
be considered as a discrete, or ``quantized'', analogue
of minimally embedding a compact (2 dimensional) surface into a
higher dimensional unit sphere (cp. \cite {2}, eq. (30)),
has a very rich spectrum of solutions-\\ some of which will now be
given:

I) Let ${\buildrel \rightarrow \over   N}$ be a basis
of a representation of some Lie-algebra $\cal G$, satisfying

\begin{equation}
[N_a,N_b]=if_{abc} N_c,~~a,b,c=1,..._{\buildrel d \over \sim },
~~~~~~f_{abc}f_{cba'}=-\lambda \delta_{aa'}
\end{equation}

I') Note that the Ansatz 
\begin{equation}
X_a(t)=X(t)\otimes N_a
\end{equation}
would reduce (1) to the (solvable) matrix model equations

\begin{equation}
{\ddot X}+X^3=0
\end{equation}
with (2) being satisfied provided $[X(0),{\dot X(0)} \bigr ]=0$.

II) Suppose $U,V~\in~{\cal A}$ satisfy 

\begin{equation}
VU=\omega UV
\end{equation}
with $\omega=e^{4\pi i\Lambda}$, $\Lambda~\in~{\bf R};$

\begin{equation}
{\buildrel \rightarrow \over  M}={1\over \sqrt {2}}\bigl (
{U+U^{-1}\over 2},{U-U^{-1}\over 2i},{V+V^{-1}\over 2},{V-V^{-1}\over
2i},
0...0 \bigr )
\end{equation}
will then satisfy (4) with 
\begin{equation}
\lambda=2 sin^2(2\pi \Lambda )
\end{equation}
--as for the above choice of ${\buildrel \rightarrow \over  M}$ one
has 
\begin{equation}
\sum_j \bigl [[\cdot ,M_j],M_j\bigr ]={1\over 2}\bigl ( 
\bigl [[\cdot ,U],U^{-1}\bigr ]+\bigl [[\cdot ,V],V^{-1}\bigr ] \bigr ),
\nonumber
\end{equation}
which is proportional to the ``quantum torus Laplace operator''
\begin{equation}
\Delta_{\Lambda}:={-1\over 16 \pi^2 \Lambda^2}\bigl ( 
\bigl [[\cdot ,U],U^{-1}\bigr ]+\bigl [[\cdot ,V],V^{-1}\bigr ] \bigr ),
\end{equation}

\begin{equation}
\Delta_{\Lambda}(U^{m_1}V^{m_2})=-{sin^2(2\pi \Lambda m_1)+
sin^2(2\pi \Lambda m_2) \over 4\pi^2 \Lambda^2 }(U^{m_1}V^{m_2}),
\end{equation}
whose eigenvalues for the 4 different components of 
$\buildrel \rightarrow \over  M$ are identical.
For rational $\Lambda $, $\Lambda={M\over N}$, $U$ and $V$
may be taken to be finite-dimensional  matrices.
The $N \rightarrow \infty$ limit of the solutions to the
corresponding matrix equations (1) then either leads to solutions 
of the continuous membrane equations (cp. \cite {1}, \cite {2}),
(or choosing $M=M(N)$ such that ${M\over N}
\rightarrow \Lambda_{\infty}\not=0$, see e.g. p. 53 of \cite {5}) 
to solutions of the ``star- product
membrane equations'' \cite {6}.

In the former case, it is known that the analogue of (4) has solutions
describing minimal embeddings into $S^{3}$ of 2-dimensional
surfaces of arbitrary genus (cp. \cite {2}). The continuous limit of (11)
and $(7)_{SO(3)}$, e.g.,  

\begin{equation}
{\buildrel \rightarrow \over  m}={1\over \sqrt {2}} (cos\varphi_1,
sin\varphi_1,cos\varphi_2,sin\varphi_2,0...0),
\end{equation}

\begin{equation}
{\buildrel \rightarrow \over  m}=(sin\theta cos\varphi,sin\theta sin\varphi,
cos\theta, 0...0),
\end{equation}
(note that just as ${\buildrel \rightarrow \over  m}^2=1$,
(11) also implies ${\buildrel \rightarrow \over  M}^2={\bf 1}$,
and in the case  of (7) irreducible representations could be normalized to
have ${\buildrel \rightarrow \over  M}^2=1$) satisfy

\begin{equation}
\lbrace \lbrace m_i,m_j\rbrace m_j\rbrace=2m_i,
\end{equation}
where $\lbrace~ ,~\rbrace$ denotes the Poisson brackets 
for functions on  $T^2$, resp. $S^2$. 

III) Let

\begin{eqnarray}
M_1={1\over 4}(S_{kl}+S_{-k-l}+S_{-kl}+S_{k-l}),~~~~
M_2={-i\over 4}(S_{kl}-S_{-k-l}-S_{-kl}+S_{k-l}),
\nonumber \\
M_3={-i\over 4}(S_{kl}-S_{-k-l}+S_{-kl}-S_{k-l}),~~~~
M_4={-1\over 4}(S_{kl}+S_{-k-l}-S_{-kl}-S_{k-l}),
\end{eqnarray}
$M_{i>4}=0$, where 
\begin{equation}
S_{\buildrel \rightarrow \over  m}:=\omega^{{1\over 2}m_1m_2}U^{m_1}
V^{m_2},
\end{equation}
${\buildrel \rightarrow \over  m}=(m_1,m_2)~\in~{\bf Z}^2$,
implying 

\begin{equation}
S_{\buildrel \rightarrow \over  m}S_{\buildrel \rightarrow \over  n}
=\omega^{-{1\over 2}(m_1m_2-m_2m_1)}S_{{\buildrel \rightarrow \over
m}+{\buildrel \rightarrow \over  n}},
\end{equation}

\begin{eqnarray}
S_{\buildrel \rightarrow \over  n}S_{\buildrel \rightarrow \over  m}
S_{-{\buildrel \rightarrow \over  n}}
=\omega^{({\buildrel \rightarrow \over  m}  \times 
{\buildrel \rightarrow \over  n})} 
S_{\buildrel \rightarrow \over
m}.
\nonumber
\end{eqnarray}
Using
\begin{equation}
\sum_{i=1}^{4}\bigl [[X,M_j],M_j\bigr ]={1\over 2}\bigl (   
\bigl [[X,S_{kl}],S_{-k-l} \bigr ]+\bigl [[X,S_{-kl}],S_{k-l} \bigr ]
\bigr ).
\nonumber
\end{equation}
and
\begin{equation}
\bigl [[S_{\buildrel \rightarrow \over  m},S_{\buildrel \rightarrow
\over  n}],
S_{-{\buildrel \rightarrow \over  n}}\bigr ]=4sin^2\bigl(
2\pi \Lambda({\buildrel \rightarrow \over  m}\times 
{\buildrel \rightarrow \over  n} )\bigr)S_{\buildrel \rightarrow \over  m}
\end{equation}

one finds that (18) satisfies (4), with 
\begin{equation}
\lambda=2 sin^2\bigl (2\pi \Lambda(2kl)\bigr )
\end{equation}
and ${\buildrel \rightarrow \over  M}^2=S_{00}\equiv {\bf 1}$. (Note that,
as long as $\Lambda$ is left arbitrary, the $kl$ degree
of freedom is somewhat redundant, and without loss of generality
one could have restricted to $k=1=l$, as was done in II).
Also note that the continuous ($\Lambda \rightarrow 0$) limit
of (19) gives minimal embedding into $S^3$ of two-dimensional surface
with self-intersections, described by

\begin{equation}
{\buildrel \rightarrow \over  m}=(cosk\varphi_1 cosl\varphi_2,
sink\varphi_1 cosl\varphi_2, cosk\varphi_1 sinl\varphi_2,
sink\varphi_1 
sinl\varphi_2,0,...0)
\end{equation}

{\bf Acknowledgement}

I would like to thank the Physics Departments of ETH Z{\"u}rich 
and Rutgers University for their hospitality.

\end{document}